\documentstyle[11pt,newpasp,twoside,epsf]{article}
\def\edcomment#1{\iffalse\marginpar{\raggedright\sl#1\/}\else\relax\fi}
\reversemarginpar
\newcounter{Rco}
\newcommand{\Ionst}[1]{\setcounter{Rco}{#1}\Roman{Rco}}
\newcommand{\Ion}[2]{\mbox{#1\ {\scriptsize\Ionst{#2}}}}

\newcommand{\logg}{\mbox{$\log g$}}
\newcommand{\loggw}[1]{\mbox{$\log g = #1$}}
\newcommand{\Teff}{\mbox{$T_{\mathrm eff}$}}
\newcommand{\Teffw}[1]{\mbox{$T_{\mathrm eff} = #1 \mathrm{kK}$}}

\begin{document}
\title{Calculation of Synthetic Ionizing Spectra for Planetary Nebulae}
\author{Thomas Rauch}
\affil{Institut f\"ur Astronomie und Astrophysik, Universit\"at T\"ubingen,
Sand 1, D-72076 T\"ubingen, Germany}

\begin{abstract}
We present a new grid (solar and halo abundance ratios) of state-of-the-art
fully line-blanketed NLTE model atmospheres which covers the parameter range
of central stars of planetary nebulae. The grid is available at the WWW. 
\end{abstract}

During their evolution, central stars of planetary nebulae (CSPN) can reach
extremely high effective temperatures (more than 100\,kK). Since NLTE effects
are particularly important for hot stars, the use of reliable NLTE stellar
model atmosphere fluxes is required for an adequate spectral analysis of these
stars.

The precise analysis of properties of planetary nebulae is strongly dependent
on the ionizing spectrum: Observations as well as NLTE model atmosphere
calculations have shown that spectra of their exciting stars neither have
something to do with black-body spectra nor can be modeled sufficiently well
with ``standard'' NLTE atmosphere models which are composed out of hydrogen
and helium only: Strong differences between synthetic spectra from these
compared to the observed spectra at energies higher than 54\,eV (\Ion{He}{2}
ground state) can be ascribed to the neglected metal-line blanketing.

For a reliable calculation of the emergent stellar flux as ionizing spectrum
for planetary nebulae the consideration of opacities from all elements from
hydrogen up to the iron-group is required (Armsdorfer et al.\,2002).

The accelerated lambda iteration (ALI) method represents a powerful tool to
calculate metal-line blanketed atmospheres with more than 300 atomic levels
treated consistently in NLTE. Thus, together with recent atomic data from the
Opacity Project and Kurucz's line lists, we can calculate realistic
atmospheres with millions of lines included.
\vspace{5mm}

\noindent
{\bf NLTE Model Atmospheres}
\vspace{5mm}

\noindent
Our model atmospheres are calculated using the NLTE code {\sc PRO2} (Werner
1986, Werner \& Dreizler 1999). The models are plane-parallel and in
hydrostatic and radiative equilibrium. More than 300 levels can be treated in
NLTE with more than 1\,000 line transitions considered in detail. Millions of
lines of the iron-group elements tabulated in Kurucz (1996) and data from the
Opacity Project (Seaton et al.\,1994) are accounted for
using an opacity sampling method (Cross-Section Creation Package CSC, Deetjen 1999,
http://astro.uni-tuebingen.de/\raisebox{1mm}{$\sim$}deetjen/csc.html).

Impacts of the light metals Li -- Ca are
investigated by Rauch (1997). The influence of the iron-group elements is
shown by Dreizler \& Werner (1993) and Deetjen et al.\,(1999).

\begin{figure}
\plotone{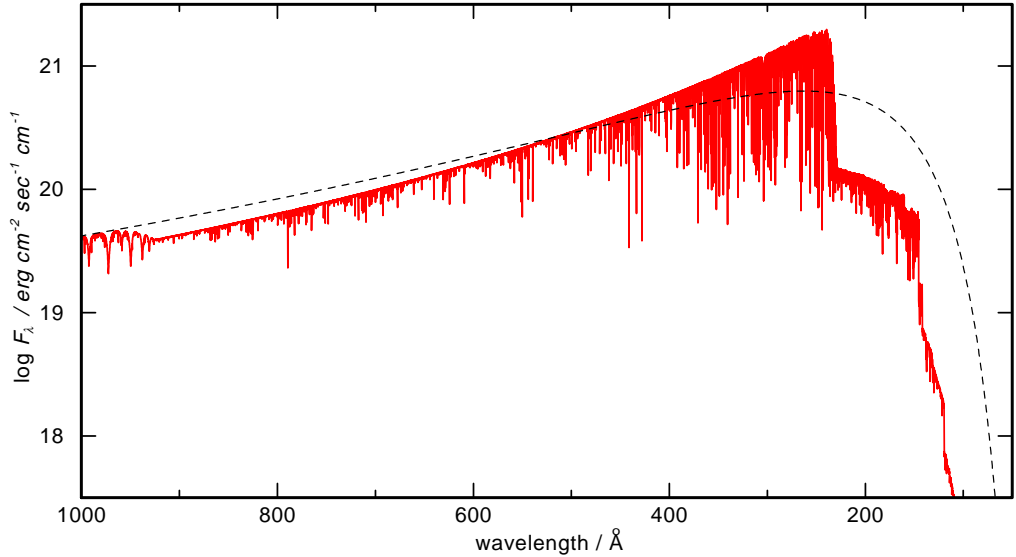}
\caption{Comparison of a synthetic NLTE-model atmosphere flux
         (\Teffw{110}, 
         \loggw{7}\ (cgs), 
         solar abundances, elements H - Ni) with a
         black body flux (same temperature)}
\end{figure} 
\vspace{4mm}

\noindent
{\bf NLTE Model Fluxes on the WWW} 
\vspace{2mm}

\noindent
A grid of H -- Ca (\Teff\ = 50 -- 1\,000\,kK, \logg\ = 5 -- 9 (cgs), solar and halo abundances)
and H -- Ni (\Teff\ up to 190\,kK) model atmosphere fluxes is available at
http://astro.uni-tuebingen.de/\raisebox{1mm}{$\sim$}rauch.

\acknowledgments
This research was supported by the DLR under grant 50\,OR\,9705\,5.


\begin{references}

\reference  Armsdorfer\,B., Kimeswenger\,S., Rauch\,T. 2002, RevMexAA in press
\reference  Deetjen\,J.L. 1999, diploma thesis, University T\"ubingen
\reference  Deetjen\,J.L., Dreizler\,S., Rauch\,T., Werner\,K. 1999,
            in: White Dwarfs, ed.\ J.-E.Solheim \& E.G.\,Mei$\breve{\rm s}$tas, The ASP Conference Series Vol.\,169, p.\,475
\reference  Dreizler\,S., Werner\,K.\,1993, \aap\ 278, 199
\reference  Kurucz\,R.L.\,1996, IAU Symp.\,176, Kluwer, Dordrecht, p.\,52
\reference  Rauch\,T. 1997, \aap\ 320, 237
\reference  Seaton M.J., Yu Yan, Mihalas D., Pradhan A.K.\,1994, MNRAS 266, 805
\reference  Werner\,K. 1986 \aap\ 161, 177
\reference  Werner\,K., Dreizler\,S. 1999, Journal of Comp.\,and Appl.\,Mathematics 109, 65
\end{references}
\end{document}